\documentclass[twocolumn]{aastex631}

\usepackage{listings} 
\newcommand{\ysepz}{{\tt YSE-PZ}}

\submitjournal{PASP}

\shorttitle{YSE-PZ}
\shortauthors{Coulter et. al.}

\graphicspath{{./}{figures/}}
 \NewPageAfterKeywords
 
\begin{document}

\title{YSE-PZ: A Transient Survey Management Platform that Empowers the Human-in-the-Loop}

\correspondingauthor{D.~A.~Coulter}
\email{dcoulter@ucsc.edu}

\author[0000-0003-4263-2228]{D.~A.~Coulter}
\affiliation{Department of Astronomy and Astrophysics, University of California, Santa Cruz, CA 95064, USA}

\author[0000-0002-6230-0151]{D.~O.~Jones}
\affiliation{Gemini Observatory, NSF's NOIRLab, 670 N. A'ohoku Place, Hilo, HI 96720, USA}

\author[0000-0002-1052-6749]{P.~McGill}
\affiliation{Department of Astronomy and Astrophysics, University of California, Santa Cruz, CA 95064, USA}

\author[0000-0002-2445-5275]{R.~J.~Foley}
\affiliation{Department of Astronomy and Astrophysics, University of California, Santa Cruz, CA 95064, USA}

\author[0000-0002-6298-1663]{P.~D.~Aleo}
\affiliation{Department of Astronomy, University of Illinois at Urbana-Champaign, 1002 W. Green St., IL 61801, USA}
\affiliation{Center for AstroPhysical Surveys, National Center for Supercomputing Applications, Urbana, IL, 61801, USA}

\author[0000-0003-0416-9818]{M.~J.~Bustamante-Rosell}
\affiliation{Department of Astronomy and Astrophysics, University of California, Santa Cruz, CA 95064, USA}

\author[0000-0003-0038-5468]{D.~Chatterjee}
\affiliation{Department of Astronomy, University of Illinois at Urbana-Champaign, 1002 W. Green St., IL 61801, USA}
\affiliation{Center for AstroPhysical Surveys, National Center for Supercomputing Applications, Urbana, IL, 61801, USA}
\affiliation{LIGO Laboratory and Kavli Institute for Astrophysics and Space Research, Massachusetts Institute of Technology, 185 Albany St, Cambridge, MA 02139, USA}

\author[0000-0002-5680-4660]{K.~W.~Davis}
\affiliation{Department of Astronomy and Astrophysics, University of California, Santa Cruz, CA 95064, USA}

\author[0000-0001-9749-4200]{C.~Dickinson}
\affiliation{Department of Astronomy and Astrophysics, University of California, Santa Cruz, CA 95064, USA}

\author[0000-0003-2348-483X]{A.~Engel}
\affiliation{Department of Astronomy, University of Illinois at Urbana-Champaign, 1002 W. Green St., IL 61801, USA}

\author[0000-0003-4906-8447]{A.~Gagliano}
\affiliation{Department of Astronomy, University of Illinois at Urbana-Champaign, 1002 W. Green St., IL 61801, USA}
\affiliation{Center for AstroPhysical Surveys, National Center for Supercomputing Applications, Urbana, IL, 61801, USA}

\author[0000-0002-3934-2644]{W.~V.~Jacobson-Galán}
\affiliation{Department of Astronomy, University of California, Berkeley, CA 94720, USA}

\author[0000-0002-5740-7747]{C.~D.~Kilpatrick}
\affiliation{Center for Interdisciplinary Exploration and Research in Astrophysics (CIERA) and Department of Physics and Astronomy,
Northwestern University, Evanston, IL 60208, USA}
\affiliation{Department of Astronomy and Astrophysics, University of California, Santa Cruz, CA 95064, USA}

\author[0009-0004-7605-8484]{J.~Kutcka}
\affiliation{Department of Astronomy and Astrophysics, University of California, Santa Cruz, CA 95064, USA}

\author[0009-0004-3242-282X]{X.~K.~Le~Saux}
\affiliation{Department of Astronomy and Astrophysics, University of California, Santa Cruz, CA 95064, USA}

\author[0000-0001-8415-6720]{Y.-C.~Pan}
\affiliation{Graduate Institute of Astronomy, National Central University, 300 Zhongda Road, Zhongli, Taoyuan 32001, Taiwan}

\author[0000-0003-4131-1676]{P.~J.~Qui\~{n}onez}
\affiliation{Department of Physics and Astronomy, Sonoma State University, Rohnert Park, CA 94928, USA}

\author[0000-0002-7559-315X]{C.~Rojas-Bravo}
\affiliation{Department of Astronomy and Astrophysics, University of California, Santa Cruz, CA 95064, USA}

\author[0000-0003-2445-3891]{M.~R.~Siebert}
\affiliation{Space Telescope Science Institute, Baltimore, MD 21218, USA}

\author[0000-0002-5748-4558]{K.~Taggart}
\affiliation{Department of Astronomy and Astrophysics, University of California, Santa Cruz, CA 95064, USA}

\author[0000-0002-1481-4676]{S.~Tinyanont}
\affiliation{Department of Astronomy and Astrophysics, University of California, Santa Cruz, CA 95064, USA}
\affiliation{National Astronomical Research Institute of Thailand, 260  Moo 4, Donkaew,  Maerim, Chiang Mai, 50180, Thailand}

\author[0000-0001-5233-6989]{Q.~Wang}
\affiliation{Physics and Astronomy Department, Johns Hopkins University, Baltimore, MD 21218, USA}

\begin{abstract}

The modern study of astrophysical transients has been transformed by an exponentially growing volume of data. Within the last decade, the transient discovery rate has increased by a factor of $\sim$20, with associated survey data, archival data, and metadata also increasing with the number of discoveries. To manage the data at this increased rate, we require new tools. Here we present \ysepz, a transient survey management platform that ingests multiple live streams of transient discovery alerts, identifies the host galaxies of those transients, downloads coincident archival data, and retrieves photometry and spectra from ongoing surveys. \ysepz\ also presents a user with a range of tools to make and support timely and informed transient follow-up decisions. Those subsequent observations enhance transient science and can reveal physics only accessible with rapid follow-up observations. Rather than automating out human interaction, \ysepz\ focuses on accelerating and enhancing human decision making, a role we describe as empowering the human-in-the-loop. Finally, \ysepz\ is built to be flexibly used and deployed; \ysepz\ can support multiple, simultaneous, and independent transient collaborations through group-level data permissions, allowing a user to view the data associated with the union of all groups in which they are a member. \ysepz\ can be used as a local instance installed via Docker or deployed as a service hosted in the cloud. We provide \ysepz\ as an open-source tool for the community.

\end{abstract}


\section{Introduction}

Time-domain astronomy is experiencing an exponentially growing rate of astrophysical transient discoveries, with 24,634 transients reported in  2021\footnote{\url{https://www.wis-tns.org/stats-maps}} compared to only 909 in 2011\footnote{\url{https://www.rochesterastronomy.org/sn2011/snstats.html}}, a 27-fold increase.  The rising discovery rate is driven by the transition from narrow-field galaxy-targeted surveys \citep[e.g., the Lick Observatory Supernova Search;][]{Filippenko2001} to wide-field time-domain surveys, including the All-Sky Automated Survey for Supernovae \citep[ASAS-SN;][]{Shappee2014}, the Asteroid Terrestrial-impact Last Alert System \citep[ATLAS;][]{Tonry2018}, the Catalina Real-Time Transient Survey \citep[CSS;][]{Drake2009}, the Gaia Photometric Science Alerts \citep{Hodgkin2021}, the Mobile Astronomical System of Telescope-Robots \citep[MASTER;][]{Lipunov10}, the Panoramic Survey Telescope and Rapid Response (Pan-STARRS) Survey for Transients \citep[PSST;][]{Huber15}, the Palomar Transient Factory \citep[PTF;][]{Law2009}, the Nearby Supernova Factory \citep[SNfactory;][]{Aldering2002}, the Young Supernova Experiment \citep[YSE;][]{Jones21:yse}, and the Zwicky Transient Facility \citep[ZTF;][]{Bellm19}. The discovery rate is expected to further increase by an additional order of magnitude with the start of survey operations for the Vera Rubin Observatory's Legacy Survey of Space and Time \citep[LSST;][]{LSSTScienceCollaboration2009}.

The corresponding continually expanding volume of data introduces new challenges for data management, transient triage, and follow-up decisions. On average, $\sim$50 new transients are reported to the International Astronomical Union (IAU) every day, with several times more {\it potential} transients identified in survey data.  Without efficient ways to sift data streams to find targets of interest, we risk missing novel transient events or failing to discover them in time to obtain follow-up observations before they have changed or faded.  Furthermore, collating data from multiple transient surveys and extant archives is essential to have the most complete dataset for making decisions in real time.

For supernovae (SNe), follow-up observations obtained within the first hours to days after explosion are particularly critical since they probe the outermost layers of a SN's ejecta and its progenitor star \citep[e.g.,][]{Soderberg2008, Modjaz2009, Ofek2010, Bloom2012, Tinyanont2022}, illuminate close-in circumstellar material before it is overrun \citep[e.g.,][]{Sternberg2011, Gal-Yam14, Jacobson-Galan20, Jacobson-Galan2022, Terreran2022}, and reveal details of the progenitor system \citep[e.g.,][]{Marion16, Hosseinzadeh17, Dimitriadis19, Shappee19, Miller20}.  While wide-field high-cadence observations are critical to discover SNe at early times, additional tools are necessary to {\it identify} these SNe before this phase has passed.

Methods of managing transient data have evolved from transient survey websites in the 1990s and early 2000s, such as the Rochester Astronomy Supernova webpage\footnote{\url{https://www.rochesterastronomy.org/supernova.html}.} that started in 1996.  This website collated information about every SN, including its location, brightness, and host galaxy.  Many SNe had finding charts, providing critical information not transmitted through International Astronomical Union (IAU) Circulars and Central Bureau Electronic Telegrams \citep[CBETs;][]{Green2002}, which at the time were the primary way for professional astronomers to communicate about SNe in (near) real time.  As IAU reporting diminished throughout the early 2000s, the Rochester webpage became the de facto database for all transients until the IAU system was overhauled and the new Transient Name Server (TNS)\footnote{\url{https://www.wis-tns.org/}.} re-engaged the community in 2016.

Internal webpages for high-redshift SN cosmology surveys including Equation of State: SupErNovae trace Cosmic Expansion \citep[ESSENCE;][]{Smith2002}, SDSS-II \citep{Frieman2008}, SNFactory \citep{Aldering2002}, and the Supernova Legacy Survey \citep[SNLS;][]{Astier2006} also tracked SN discoveries from their specific surveys and included important information about the sources, such as brightness and classification.  Over time, these systems became more sophisticated. Several systems began to splinter into ``search'' services and ``target and observation managers'' (TOMs). The former would be a database of potential transients from the survey \citep[e.g.,][]{Goldstein2015}, while the latter would consist of ``promoted'' transients of interest that could be monitored. For several surveys, these internal tools were significantly more powerful than the Rochester webpage or the CBETs, causing further fragmentation.

In recent years, the community has tackled the challenge of effectively and efficiently acting on modern transient data streams by breaking the problem into two complementary layers: so-called data ``brokers,'' which broadly replicate the search features of previous surveys but with several enhancements; and continually improved TOMs. Data brokers parse the data from raw transient alert streams by filtering on criteria such as those that reduce false positive detections or indicate likely SNe.  Many brokers have focused on the ZTF alert stream, which has $\sim$1 million alerts a night, including MARS \citep{Brown2013}, ANTARES \citep{Saha2014}, Lasair \citep{Smith2018}, AMPEL \citep{Nordin2019}, Fink \citep{Moller2021}, ALeRCE \citep{Forster2021}, and Pitt-Google\footnote{\url{https://github.com/mwvgroup/Pitt-Google-Broker}.}.  Nearly all of these brokers will also be tasked with serving the Rubin Observatory alert stream --- which will provide $\sim$10 million alerts per night --- as it begins operation in the mid 2020s.  Some of these brokers include features such as value-added galaxy catalog cross-matching (e.g., via {\tt GHOST}; \citealt{Gagliano21}). Several also give the user flexibility to define their own transient filtering criteria and execute queries on a flexibly-defined set of data attributes.

After transient data have been parsed by a broker, a TOM will store the pre-processed alert packages and facilitate follow-up observations or identify samples for subsequent scientific analyses. Some recently developed TOMs include the Palomar Transient Factory (PTF) Follow-up Marshal \citep{Rau2009}, the RoboNet Microlensing System \citep{Tsapras2009}, the PESSTO Marshall \citep{Smartt15}, the Supernova Exchange \citep[SNeX;][]{Howell17}, the TOM-toolkit \cite{Street2018}, the Global Relay of Observatories Watching Transients Happen (GROWTH) Marshall \citep{Kasliwal19}, the Transient Science Server \citep{Smith2020}, the SNAD ZTF object viewer \citep{Malanchev2022}, and the NASA Exoplanet Follow-Up Portal (ExoFOP)\footnote{\url{https://exofop.ipac.caltech.edu/tess/}}.

These tools are usually designed for a single survey and purpose, with common limitations including a focus on specific data sets or surveys, the lack of access to the public, and a closed source codebase. The exception is the TOM-toolkit \citep{Street2018}, which is an open-source{\footnote{\url{https://github.com/TOMToolkit}.}.} extensible and customizable TOM framework, but --- by design --- requires significant customization by the user. Despite the rich combination of brokers and TOMs currently available, it remains challenging to leverage data into effective decision making. Automated tools that retrieve metadata, fit models, or classify transients often need human intervention to robustly identify transients of interest and select scientifically interesting samples of transients. 

Here, we address these ongoing challenges with \ysepz, a transient survey management platform. \ysepz\ combines traditional TOM functions, e.g., data collation and resource management, with a data broker-like query and filtering system to empower human-in-the-loop decision making.  It is open-source, publicly available software that synthesizes astronomical data from existing public transient survey and combines these data with private data, user-uploaded data, archival metadata, and output from value-added services. \ysepz\ synthesizes and displays these distinct data in an easy-to-digest format, provides a variety of query and filtering tools that supports filtering these data streams into scientific samples of interest, and enhances real-time decision-making. Finally, \ysepz\ includes a framework for planning and executing follow-up observations, allowing transients to be managed from discovery through to analysis of followup data.

In Section~\ref{sec:overview} we give an overview of the \ysepz\ platform. In Section~\ref{sec:workflow} we discuss our generic transient workflow, and in Section~\ref{sec:casestudies} we share use cases that illustrate how \ysepz\ supports a diverse set of user and science requirements. We discuss YSE-PZ in the context of a broader open science ecosystem in Section~\ref{sec:discussion}.  We conclude in Section~\ref{sec:conclusions}.

\section{YSE-PZ}
\label{sec:overview}

\begin{figure*}[t!]
    \centering
    \includegraphics[width=1.0\textwidth]{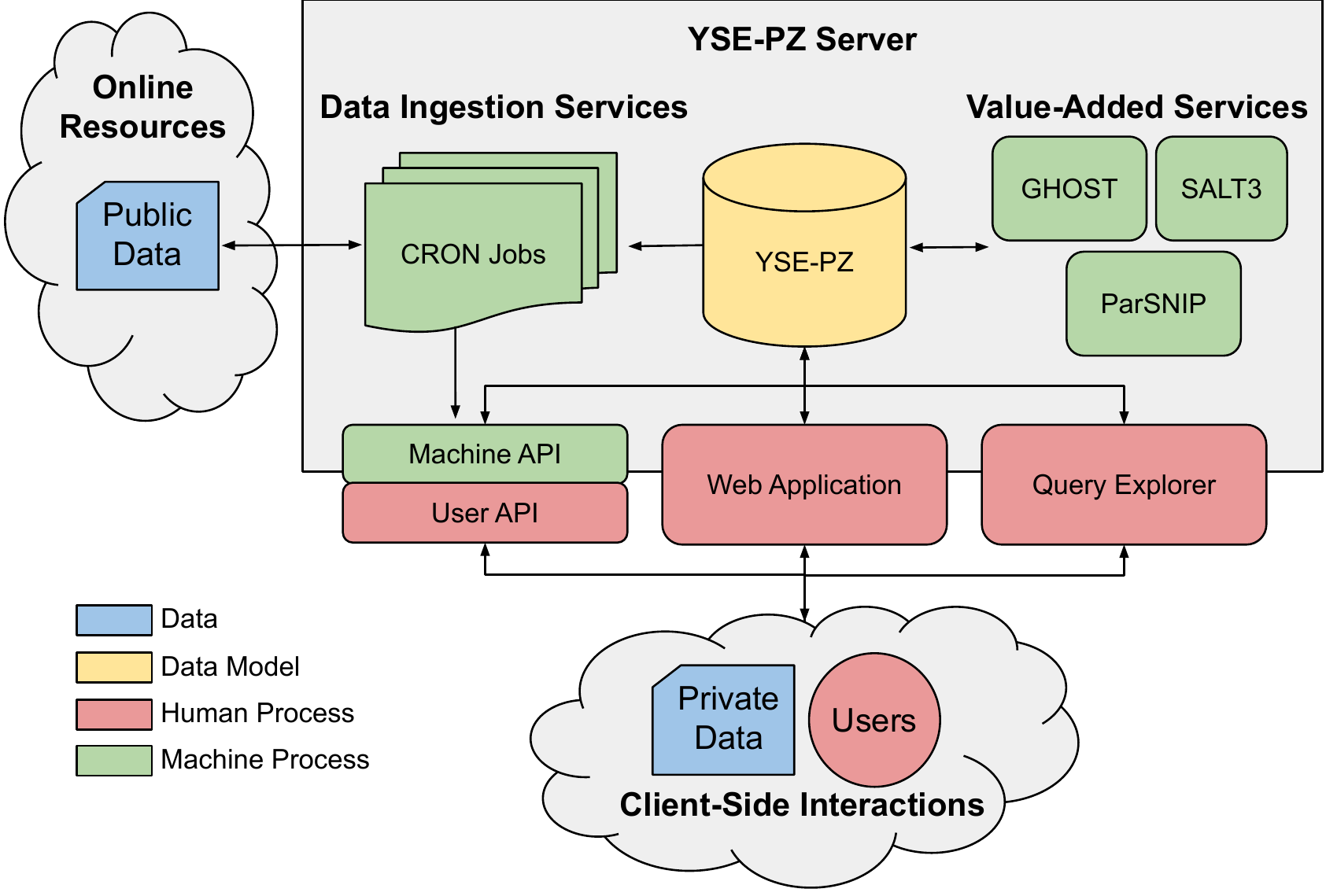}
    \caption{Architectural schematic of \ysepz. Arrows show the flow of data between entities. All entities within the rectangle are housed on the UCSC Transients Team's research server. Users of \ysepz\ are associated with a survey and interact with the application via the API, web user interface (UI), and the Query Explorer. CRON Jobs, which are run server-side, request data from external public services. These data are posted back to the application via the API. Value-added services including host-galaxy association via {\tt GHOST} \citep{Gagliano21}, transient classification via {\tt ParSNIP} \citep{Boone2021}, and transient light-curve fitting with {\tt SALT3} \citep{Kenworthy21} are also run server-side and populate the \ysepz\ database with auxiliary transient information.}
    \label{fig:architecture}
\end{figure*}

\ysepz\ is a new transient survey management platform that takes an object-oriented approach to modeling the full workflow of transient observations. This workflow is built on \ysepz’s data model, which defines relationships between the transient, its data, and metadata.  This model also defines the types of data that \ysepz\ can store and upon which it can act. \ysepz\ is also a dynamic application; it continually ingests new transients, their data and metadata, invokes value-added services that annotate these data, and performs application-level maintenance tasks.

In this section, we will first cover \ysepz's data model (Section~\ref{datamodel}) and task management system (Section~\ref{sec:crons}), and then we will enumerate the data (Section~\ref{crondataingestion}~and~\ref{archival_data}) as well as features (Section~\ref{sec:valued-added services}) enabled by these two components. We then describe how a human-in-the-loop interacts with \ysepz\ via the query engine (Section~\ref{sec:queries}), front-end web application (Section~\ref{sec:frontend}), API (Section~\ref{sec:api}), and user groups and permissions (Section~\ref{sec:groupsandpermission}).

\subsection{Data Model}
\label{datamodel}

\ysepz\ is a Django-based web application\footnote{\url{https://www.djangoproject.com/}} employing a MySQL\footnote{\url{https://www.mysql.com/}.} backend and a Representational State Transfer (REST) compliant Application Programming Interface (API) (see Figure~\ref{fig:architecture}).  To model the objects within the transient workflow, \ysepz\ uses the Django Object-Relational Mapper (ORM) framework. The ORM allows a developer to model transient properties, behaviors, and relationships within Python code rather than creating these objects directly in SQL. This developer-friendly approach makes extending the application easier. 

\ysepz's data model is constructed to be general enough to model any transient survey data, astronomical metadata, and observational workflow. While this generality makes some queries more complex than those against a simpler data model, the advantage is that \ysepz\ can ingest any public or private astronomical data and define workflows down to the instrument configuration for follow-up requests. 

\begin{figure*}
    \centering
    \includegraphics[width=0.7\textwidth]{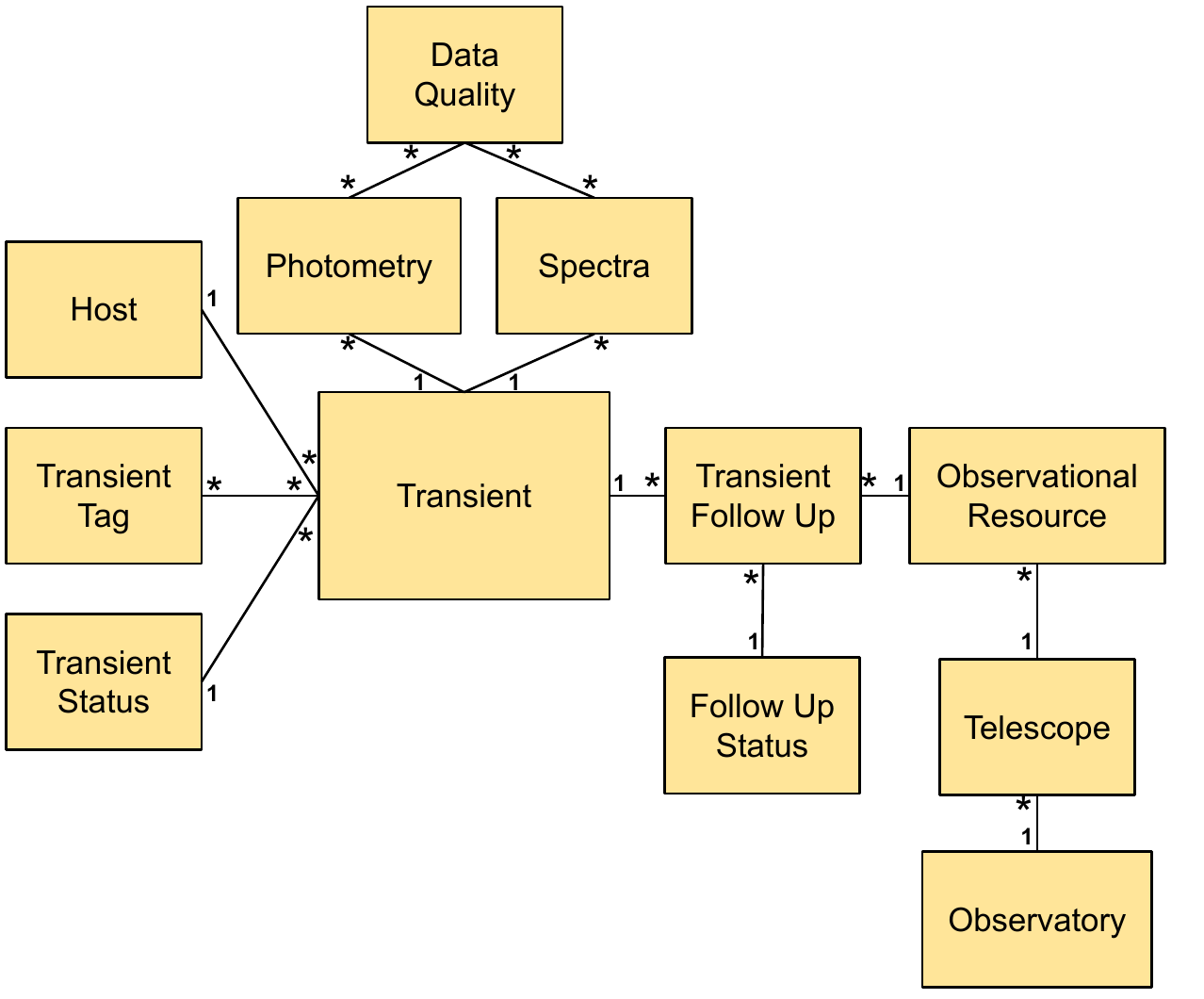}
    \caption{Simplified \ysepz\ data model. This schematic shows how a representative sample of objects within the schema are related and anchored to a transient. Lines that terminate in a ``1'' denote a singular relationship, while lines that terminate in a ``*'' denote a ``many'' relationship, e.g., {\tt Transient} has a many-to-many relationship with {\tt Transient Tag}, but a one-to-many relationship to {\tt Photometry}. See Section \ref{datamodel} for an expanded discussion.}
    \label{fig:schema}
\end{figure*}

Within \ysepz, the main objects include the transients themselves, their host galaxies, metadata, public data, private data, and auxiliary data such as observing resources and follow-up requests. \ysepz\ has 100 tables, for a complete listing of tables, see the \ysepz\ GitHub repository\footnote{\url{https://github.com/davecoulter/YSE_PZ}}. For brevity, we present a simplified science schema of \ysepz\ in Figure~\ref{fig:schema}, conceptually focused on the central \ysepz\ objects. A {\tt Transient} object is defined by its name, coordinates, and discovery date. A {\tt Transient} object is also connected to other data and metadata objects. For example, a {\tt Transient} has associated astronomical data objects, {\tt Photometry} and {\tt Spectra}, each defined by fields like flux, magnitude, bandpass, etc. A {\tt Transient} also has an associated metadata object called {\tt Transient Status} that stores how a given transient relates to a user's workflow (see Section \ref{sec:workflow}). 

In the case of {\tt Photometry} and {\tt Spectra}, a {\tt Transient} has a ``one-to-many'' relationship, i.e., one transient may have many data points, but each data point is associated with only one transient. On the other hand, some object relations are ``many-to-many'', e.g., the {\tt Transient}-{\tt Transient Tag} relationship. The {\tt Transient Tag} object holds a user-defined tag that can be applied to any transient, and denotes some interesting property (see Section~\ref{sec:queries}). One tag may apply to many transients, and a transient may have many different tags for many different properties. These relationships are denoted in Figure~\ref{fig:schema}.

\subsection{Front-End Web Application}
\label{sec:frontend}

\begin{figure*}[t!]
    \centering
    \includegraphics[width=\textwidth]{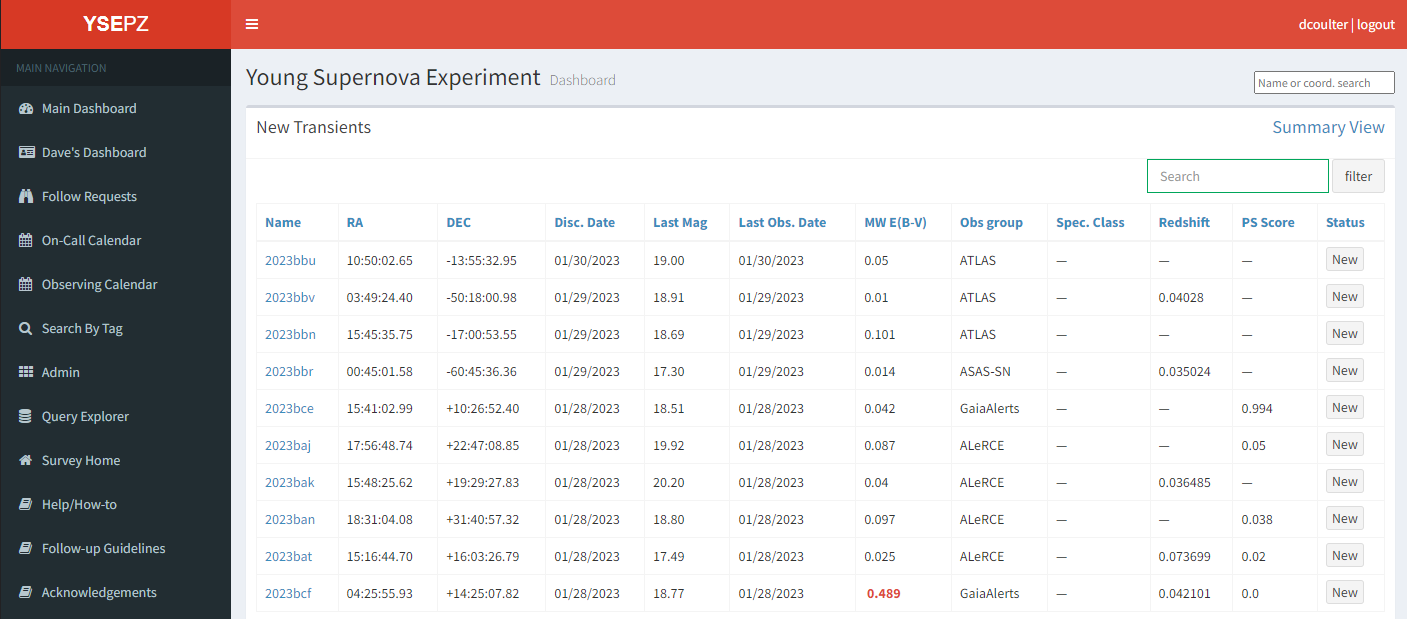}
    \caption{One of \ysepz's main dashboard tables. Transients with ``New'' status are displayed in a searchable ``New Transients'' HTML table with associated summary information for each transient (see Section \ref{sec:workflow} for a description of how the transient statuses are used). In addition to the table shown, tables for transients grouped by other statuses (``Following", ``Interesting", ``Watch", ``Finished Following", and ``Needs Template") are available further down the dashboard.  Milky Way reddening values in the ``MW E(B-V)'' column are automatically displayed in red when they are $>$0.2~mag.  The left navigation bar provides easy access to the rest of \ysepz's features and views.}  
    \label{fig:main_dashboard}
\end{figure*}

The Django ORM also powers a front-end web application in the industry-standard design pattern of the Model-View-Controller (MVC) architecture. In this paradigm, web requests are made to a controller object, which acts as a gateway that routes incoming requests to the appropriate resource, e.g., a web page or direct data download.

Based on the requested URL, user identity, and group permission, the controller will retrieve the information of interest from the database using the Django ORM and then package these data into an object called a {\tt model}. A {\tt model} is a dictionary of structured data that is rendered dynamically by an HTML template, called a {\tt view}. The final rendered HTML is what a user sees in their browser. 

\ysepz\ accomplishes most of its front-end functionality with a series of interactive dashboards that allow data to be displayed in space-efficient, paginated tables with the ability to sort on each column displayed. Upon logging into \ysepz, a user is greeted with the main dashboard, which displays newly ingested transients from TNS into a table called ``New Transients" (Figure~\ref{fig:main_dashboard}). Below the New Transients table, there are further tables for transients with different user-configurable statuses: {\it Followup Requested}, {\it Following}, {\it Interesting}, {\it Watch}, {\it Finished Following}, and {\it Needs Template} (see Section~\ref{sec:triage} for a description of these statuses). While these are the default statuses, it's important to note that a user is free to define any status they wish, as well as to customize the main page of the application to reflect their workflow preferences.

\begin{figure*}[t!]
    \centering
    \includegraphics[width=0.833\textwidth]{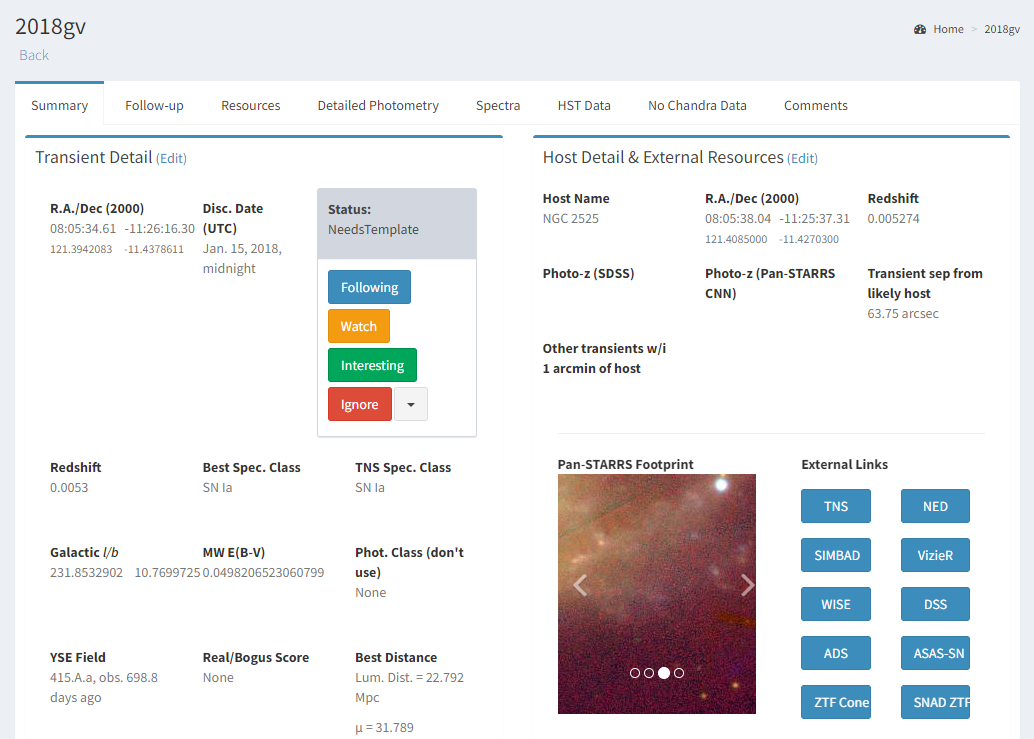}
    \includegraphics[width=0.833\textwidth]{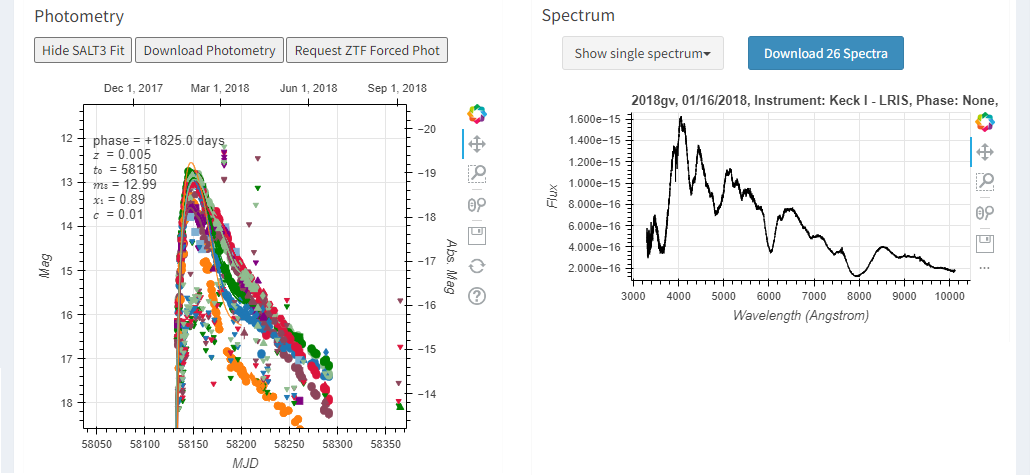}
    \caption{Subset of elements of a \ysepz\ transient detail page, which shows information related to a single transient (in this case, SN~2018gv; \citealt{Yang2020}).  The top-left quadrant enumerates the basic positional information for a transient and includes controls for changing a transient's status. The top-right quadrant focuses on host-galaxy data, including postage stamp images and links to external web services. In the bottom-left quadrant, there is an interactive Bokeh \citep{Bokeh18:bokeh} light curve plot showing photometric data for the transient. Three buttons allow the user to toggle a SALT3 \citep{Kenworthy21} model light curve fit to the photometry, download the photometric data, and request forced photometry for the transient from ZTF. The bottom-right quadrant shows an interactive plot of the spectral data for the transient, with buttons allowing users to display all spectra for the transient simultaneously or individually, and to download the data. From left to right, the tabs along the top panel include the Summary tab, which is currently selected in this Figure, tools for requesting follow-up observations for the given transient (the Follow-up tab), a list of observation resources that can be triggered for a transient with per-observatory airmass plots and a utility to generate finder charts (the Resources tab), a downloadable, tablular-formatted list of all photometry on a transient (the Detailed Photometry tab), detailed spectral information (the Spectra tab; not currently implemented), {\it Hubble Space Telescope} and {\it Chandra X-ray Observatory} archival data at the transient location, respectively, and a text area allowing users to attach free-form comments to a transient (Comments tab).}
    \label{fig:detail_page}
\end{figure*}

The front-end web application also has a left navigation bar with links to other parts of the site, including a separate web page that displays all follow-up requests made for each telescope resource, an on-call calendar which is configurable to send SMS text messages to users who want to respond to events in real time (e.g., searching for gravitational wave counterparts, see Section \ref{sec:swopesnsurvey}), an observation calendar (which displays configured observing resources in a calendar format, see Figure~\ref{fig:obs_calendar}), a search utility that can query the database for transients that have been tagged with user-defined tags (see Figure~\ref{fig:tags}), a Query Explorer\footnote{The \ysepz\ Query Explorer is built off of the {\tt Django SQL Explorer} \url{https://django-sql-explorer.readthedocs.io/en/latest/}} that allows users to flexibly create ad hoc queries against the database on any property of the data model, and an administration function exposed by Django that allows a user with administrative permissions to edit data in the database directly using a built-in, Django web form. Finally, each user has their own personal dashboard that can be configured to display transients of interest. To do this, a user writes a query to select objects using the Query Explorer, and then attaches that query to their dashboard. Users can also select from a predefined set of python-based queries for this purpose. 

Transient data can then be viewed via template-generated, individual transient detail pages (see Figure~\ref{fig:detail_page}). These transient detail pages are tab-based, and each tab provides different views of the data and different functions available to a user. On the main Summary tab, all available photometry and spectroscopy that a user is permitted to view (see Section~\ref{sec:groupsandpermission}) is plotted via an interactive {\tt Bokeh} JavaScript widget. Metadata and external links to archival data (e.g., NED and TNS) are shown at the top of the page, and archival Digitized Sky Survey (DSS), SDSS, Pan-STARRS, and DECam legacy survey images are shown in a section devoted to galaxy host data. On the Follow-up tab, there are resources for requesting follow-up for a transient, as well as forms to add new observational resources to the system (see Section~\ref{sec:followup}. A Detailed Photometry tab provides a tabular view of all transient photometry, as well as a convenient way to download this data. {\it HST} and {\it Chandra} tabs report if archival image footprints coincide with the transient position, and a Comments tab allows users to attach free-form comments to a transient.

\subsection{\ysepz's Task Management System}
\label{sec:crons}

The dynamic part of \ysepz\ is the configured Command Run On (CRON) job. These task runners are modular, configured for a single task, and interact with the API via {\tt POST}s, as well as the Django ORM. The modular nature of these CRON jobs makes extending \ysepz's functionality straightforward. In principle, CRON jobs can be run from any server and interact purely with the API to {\tt GET} or {\tt POST} any necessary data; however, \ysepz\ largely uses CRON jobs that are constructed to directly query the database on the \ysepz\ back-end for computational efficiency. These server-side CRON jobs are constructed and organized using the {\tt django-cron}\footnote{\url{https://django-cron.readthedocs.io/en/latest/}.} module, which allows users to create and configure CRON jobs using Django's application settings file, as well as providing access to the ORM's representation of \ysepz\ objects. 

\ysepz\ uses CRON jobs to continually ingest public data (see Section~\ref{crondataingestion}), retrieve archival and metadata (see Section~\ref{archival_data}), and invoke value-added services (see Section~\ref{sec:valued-added services}). CRON jobs are also used by \ysepz\ to perform various other tasks, including annotating data (e.g., flagging transients in {\it TESS} fields), scheduling YSE survey observations (see Section~\ref{sec:casestudies}), and performing daily database back-ups. For more on how \ysepz\ is used by specific collaborations, see Section~\ref{sec:casestudies}.

\subsection{Sources of Publicly Available Transient Data}
\label{crondataingestion}
\ysepz\ retrieves and stores data from the following sources via CRON jobs:

\begin{enumerate}
\item {\bf The Transient Name Server} (TNS\footnote{\url{https://www.wis-tns.org/}.}).  TNS is the official reporting service for the International Astronomical Union (IAU) that provides available photometry and spectroscopic classifications for publicly reported transients discovered by the community.  The CRON job that handles TNS data ingestion queries their API every five minutes to ingest newly discovered transients or to update previously ingested transients with new photometric, spectroscopic, and classification data. Because existing transients may be updated within TNS, this CRON job has logic to prevent transients from being duplicated\footnote{We require that any new transient be greater than 2~arcsec from the position of an existing transient {\it or} be discovered more than a year earlier/later.  While this still allows for duplicate events in rare cases with an exceptionally long reporting/discovery lag time, it generally mitigates against missing new transients at a coincident location in the same galaxy; an occurrence that could be particularly scientifically interesting.}.

\item {\bf ATLAS forced photometry} \citep{Tonry2018, Smith2020, Shingles2021}.  The Asteroid Terrestrial-impact Last Alert System (ATLAS) surveys the visible sky in the ``cyan" and ``orange" bands (``cyan" is approximately the same as $g+r$; ``orange" is approximately $r+i$) every two days. \ysepz\ automatically queries the public ATLAS forced photometry data once per day for a subset of transients of interest that we identify using a \ysepz\ query.

\item {\bf Gaia photometric science alerts} \citep{Hodgkin2021}.  Gaia observes the whole sky with a spatially heterogeneous cadence \citep[e.g.,][]{Boubert2020} but typically makes successive pairs of observations separated by 2-4 weeks in the G-band (a broad white-light filter). Gaia alerts are queried hourly.

\item {\bf PSST via the Transient Science Server}.  PSST surveys $\sim$14,000 square degrees of sky in the $i$ and $w$ (white light that is roughly $g+r+i$) bands at an irregular cadence with 1-4 return visits of each field scheduled within 15 days of the first observation.  PSST photometry is ingested every two hours.

\item {\bf{\it Swift} optical and ultraviolet photometry via the {\it Swift} quick-look data archive}\footnote{\url{https://swift.gsfc.nasa.gov/cgi-bin/sdc/ql}}.  Once per hour, we query all previously unanalyzed {\it Swift} imaging for overlap with transients in YSE-PZ and perform forced aperture photometry on these images at the locations of the transients.

\item {\bf The Young Supernova Experiment via the Transient Science Server} \citep{Smith2020}.  The Young Supernova Experiment surveys $\sim$1500 square degrees of sky at any time to a depth of $gri \simeq 21.5$~mag, and $z \simeq 20.5$ mag, with a three-day cadence.  YSE data are vetted using the Transient Science Server, and transients deemed both ``good" and ``possible" (indicating a transient that may be real) by the YSE team are ingested to \ysepz\ every 30 minutes.

\item {\bf ZTF via the ANTARES Astronomical Time-domain Event Broker} \citep[][]{Matheson2021}. ZTF is currently surveying the northern extragalactic sky on a two-day cadence, with publicly available photometry in the $gr$ bands.  We use the ANTARES client to ingest ZTF photometry for transients reported to TNS; ZTF photometry is queried upon ingestion of the TNS transients and updated twice per day for transients not flagged as ``Ignore" (see Section \ref{sec:triage} for status labels).  Forced photometry can be manually requested via the ZTF forced photometry service\footnote{\url{https://ztfweb.ipac.caltech.edu/cgi-bin/requestForcedPhotometry.cgi}} and is uploaded within approximately one hour, depending on the speed of the ZTF forced photometry server\footnote{Previously, \ysepz\ used the ``Make Alerts Really Simple" \citep[MARS, from the Las Cumbres Observatory;][]{Brown2013} broker to perform the same task, but this service has been taken offline by Las Cumbres Observatory at the end of January 2023.}.

\item Additional photometric and spectroscopic data from individual collaboration follow-up surveys currently using \ysepz\ are continually added to \ysepz\ via private scripts that {\tt POST} their data through the API. This workflow demonstrates the extensibility of the platform, as well as an example of API-only CRONs referred to in Section~\ref{sec:crons}. 

\end{enumerate}

While TNS is the primary source for transients to enter the \ysepz\ database, transients can also be directly discovered by collaborations that use \ysepz, and ingested via the API (see the YSE Case Study in Section~\ref{sec:casestudies}). In these cases, \ysepz's API has been designed to match existing transients by position, taking a union of the final objects' properties while setting the default transient name to the IAU name from TNS.

\subsection{Sources of Archival data}
\label{archival_data}

Archival data and metadata, including static-sky data and forced photometry, are retrieved for each transient using several public catalogs and image servers.  Cutout archival images at the location of each transient are displayed using image servers from SDSS \citep{York2000}, Pan-STARRS \citep{Chambers2016}, the Digitized Sky Survey via Aladin \citep{Bonnarel2000,Boch2014}, and the DESI Legacy Imaging Surveys \citep{Dey2019}. These data are not stored in the database but can be easily downloaded by the user.  Any available {\it HST}, {\it Spitzer}, and {\it Chandra} data are included on each transient detail page (see Section~\ref{sec:frontend}), and the existence of such archival data from each source is stored in the {\tt Transient} table to facilitate queries on transients with archival space-based data. Archival Pan-STARRS1 catalog data \citep{Chambers2016, Flewelling2020} are also used to determine whether a given transient coincides with a point source \citep{Tachibana2018}, indicating a likely flaring star.

\subsection{Value-added services}
\label{sec:valued-added services}

\ysepz\ uses {\tt GHOST} \citep{Gagliano21} to match each transient to its most likely host galaxy; {\tt GHOST} also  measures a photometric redshift via {\tt Eazy PhotoZ} \citep[][]{Aleo2022} for each host galaxy from a fully connected, feed-forward neural network algorithm trained on Pan-STARRS data.  \ysepz\ also includes interactive plots and plot-anlysis tools. Plotting is performed with Bokeh \citep{Bokeh18:bokeh}, and interactive {\tt SALT3} fitting is performed \citep{Kenworthy21} using {\tt sncosmo} \citep{barbary21}. These routines estimate light-curve parameters and approximate times of maximum light under the assumption that a transient is a Type Ia SN.  Finder charts for each transient can be generated automatically on request from the user, and airmass plots, target rise/set times, and moon angle information from {\tt astroplan} \citep{astroplan2018} are also included (see Section~\ref{sec:followup} for a discussion on these features).

\subsection{Queries \& Tags}
\label{sec:queries}

All data in the \ysepz\ database can be queried either through the REST API or the Query Explorer. These queries are custom written to meet a user or survey group's transient science goals. The most-used queries are those that identify recently discovered transients with rising light curves and queries that find unclassified transients in a magnitude- or volume-limited sample. For example, the Keck Infrared Transient Survey (Section \ref{sec:casestudies}), uses these queries to identify subsets of transients for follow-up observations. The results of these queries can be displayed on a user's personal dashboard page or accessed programmatically through the web application itself (see Appendix \ref{sec:example_queries} for example queries).

\begin{figure}
    \centering
    \includegraphics[width=0.49\textwidth]{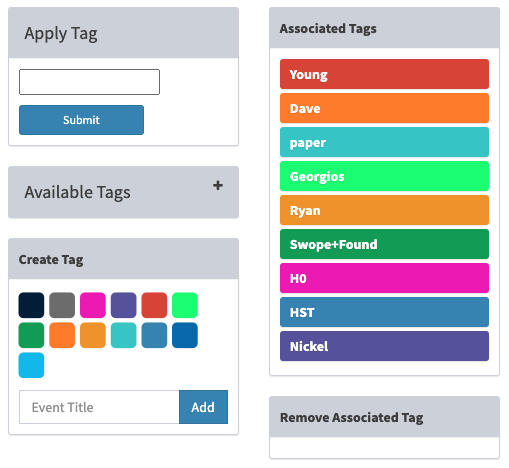}
    \caption{The \ysepz\ transient tag form, which is located on each transient detail page. This form allows a user to associate arbitrary tags with a given transient, allowing configurable and user-customized tracking.}
    \label{fig:tags}
\end{figure}

While queries provide a powerful way to interrogate the data, \ysepz\ also enables users to apply customizable tags to objects (see Figure~\ref{fig:tags}). Tags are user-defined, text strings that can be attached as metadata to transients. These tags empower users to create arbitrary groupings of transients, and provides an additional property upon which queries can be built. Tags can also be applied automatically to facilitate data triage from certain sources or science programs.

\subsection{Application Programming Interface}
\label{sec:api}

\ysepz\ contains two Application Programming Interfaces (APIs), a machine-facing API used to ingest or extract raw data from an authorized user into \ysepz's tables directly, and a user-facing API that allows users to upload data to \ysepz\ through a system of checks and flags that allow data to be created, modified, and deleted while checking for logical inconsistencies and errors like incomplete data, incorrectly formatted data, or duplication. The machine API is provided by a Django plugin module called the Django REST framework, while the user API was written by \ysepz\ developers. Both APIs are easily extensible and continue to evolve as needed to accommodate new science requirements.

\begin{figure*}[t!]
    \centering
    \includegraphics[width=0.48\textwidth]{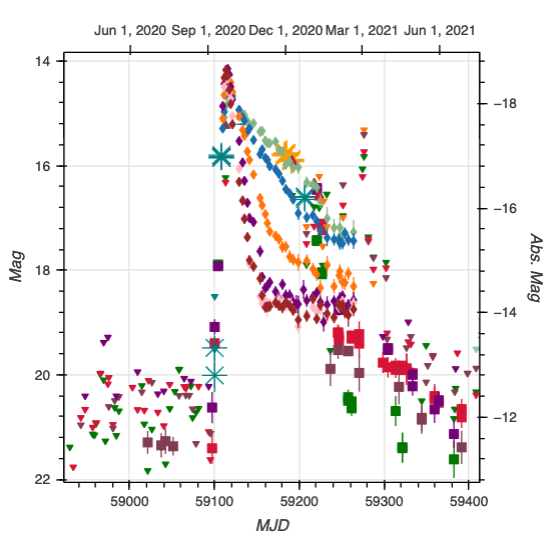}
    \includegraphics[width=0.50\textwidth]{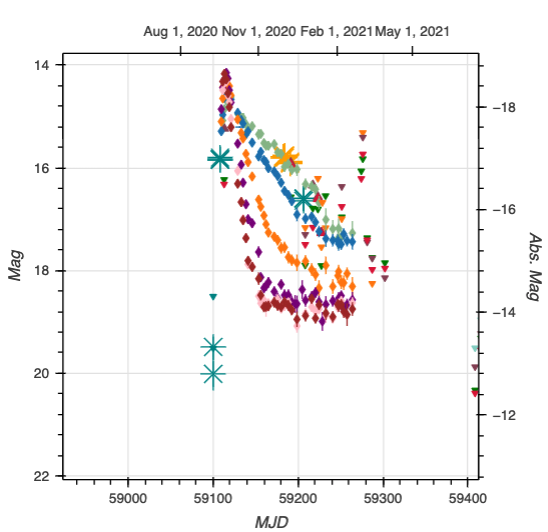}
    \caption{An example \citep[SN~2020tlf;][]{Jacobson-Galan2022} of the effect of \ysepz\ group permissions as shown through the displayed light-curve plot on an object's detail page. Detections and non-detections are displayed as different symbols including an error bar (although sometimes smaller than the plotted symbol) and downward-pointing triangles, respectively.  Left: The full light curve of a transient as seen by a user with membership in all \ysepz\ groups.  Right: A view of the same transient from the perspective of a user with membership in only a select number of \ysepz\ groups. In this case, users with access to YSE Pan-STARRS data (displayed as squares) could see the pre-explosion brightening described by \citet{Jacobson-Galan2022}, while those without access to those private data would not see this behavior.  Since there is a gap in the YSE Pan-STARRS data not long after explosion, the combination of multiple data sources were critical to follow the evolution of SN~2020tlf.}
    \label{fig:permissions}
\end{figure*}

\begin{figure}
    \centering
    \includegraphics[width=0.40\textwidth]{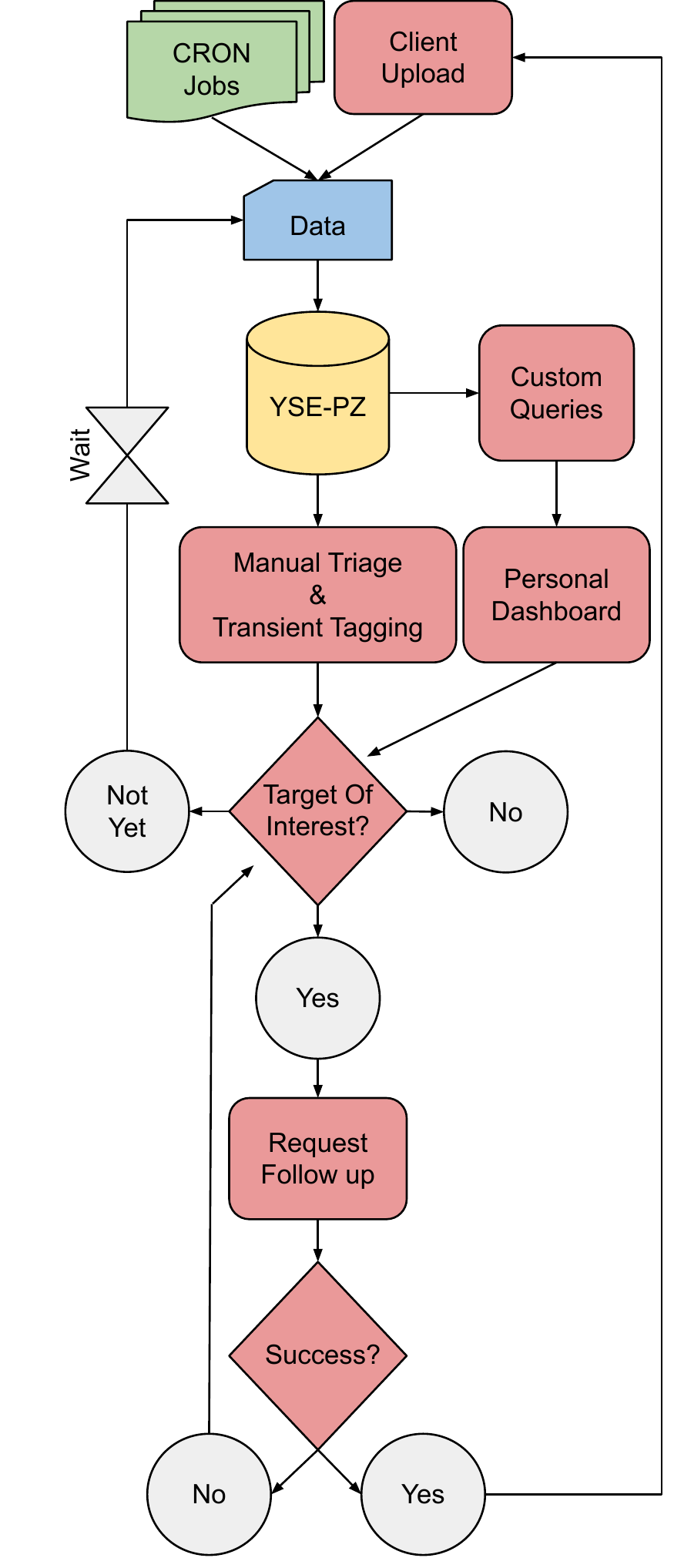}
    \caption{Data flow within \ysepz. New transients, as well as their scientific data and metadata, are ingested into \ysepz\ and then triaged by a human either through manually screening new transients, tagging individual transients, or by writing queries based on various transient properties. After identifying interesting candidates, follow-up requests can be made and subsequent data can be uploaded.}
    \label{fig:workflow}
\end{figure}

\begin{figure*}[t!]
     \centering
    \includegraphics[width=0.75\textwidth]{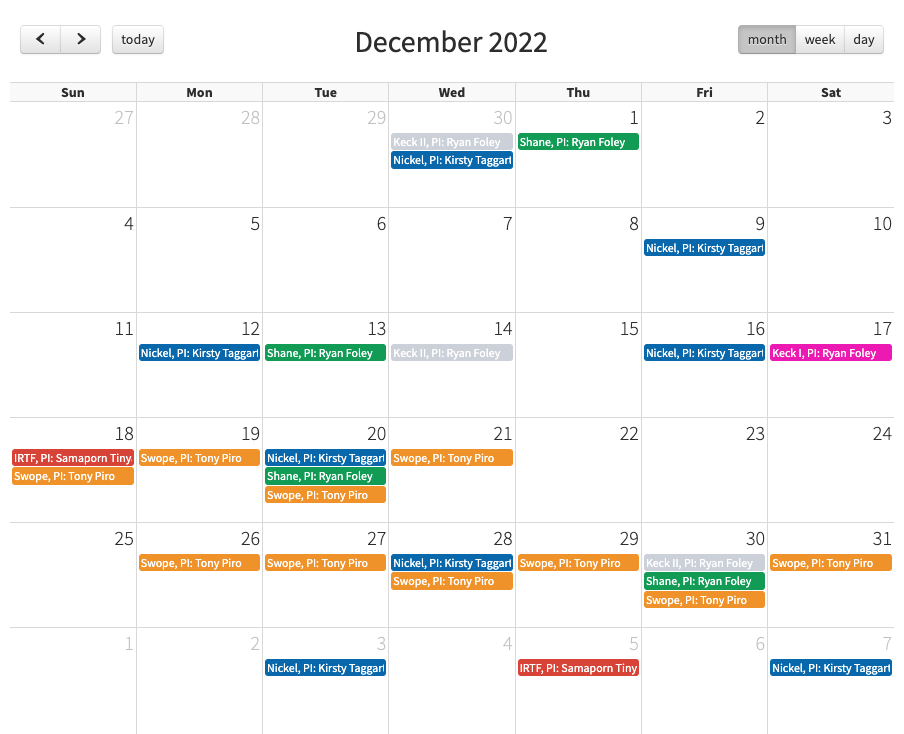}
    \caption{Observing calendar showing all scheduled resources for a given month. Both the observing asset and Principal investigator (PI) for each resource is shown.}
    \label{fig:obs_calendar}
\end{figure*}

\begin{figure*}[t!]
     \centering
    \includegraphics[width=0.75\textwidth]{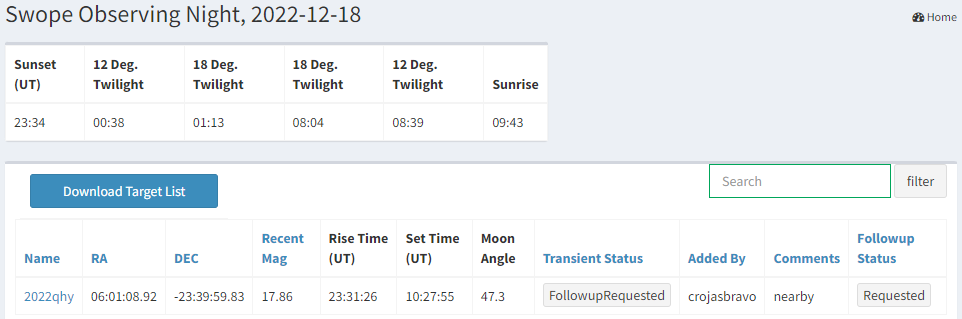}
    \caption{Partial screenshot of an observing night schedule webpage showing requested transient targets and their associated basic information and status. Users are able to download the target list from this page.}
    \label{fig:obs_night}
\end{figure*}

\subsection{User Groups and Permissions}
\label{sec:groupsandpermission}

\ysepz\ has been designed to support multiple surveys and collaborations within a single instance. This is accomplished through user-defined ``user groups'' that can be associated with astronomical data uploaded through the API. Access to these data (both download and display) is then controlled by adding or removing users to specific groups. An example of this data access control is shown in Figure~\ref{fig:permissions}.

While user groups offer reasonable control over access to private data, we recommend that if data security is a concern, a separate instance of {\ysepz}\ be deployed. In this scenario, there is no possibility of one group gaining access to another group's data on the same system. See Section \ref{sec:discussion} for a discussion on {\ysepz}\ deployment approaches.

\section{The YSE-PZ transient Lifecyle}
\label{sec:workflow}

In this section we detail how \ysepz\ is used to triage, analyze, and obtain follow-up data for transients. Figure~\ref{fig:workflow} shows an overview of the entire life cycle of an ingested transient in \ysepz. We explain each aspect of the workflow below.

\subsection{Manual Triage and Transient Tagging}
\label{sec:triage}

When transients are first ingested into \ysepz, they are given a status of {\it New}.  As described in Section~\ref{sec:frontend}, users interact with these transients through a series of dashboards (e.g., Figure~\ref{fig:main_dashboard}) with sortable, paginated tables that include redshifts, classifications, Milky Way reddening, last reported magnitudes, and discovery dates for each transient.  These provide the user an overview of the data and allow them to easily view and triage new transients.

Though the exact status categories can be altered by the user, our team has found the following status categories to be useful:

\begin{itemize}
    \item {\it New}: a transient that has just entered the system.
    \item {\it Watch}: a transient that could be potentially interesting, but hasn't risen to the level of being targeted for follow-up observations. Most transients are triaged into this category from their initial status of {\it New}.
    \item {\it Interesting}: an interesting transient that may be targeted for follow-up observations.
    \item {\it Followup Requested}: A transient for which a follow-up request has been made.
    \item {\it Following}: a transient actively being followed. At least one observation is necessary for this status to be used.
    \item {\it Needs Template}:  a transient with completed follow-up observations, but that require template observations (i.e., observations with no transient flux).
    \item {\it Followup Finished}: a transient whose dedicated follow-up observations have been completed, including template observations.
    \item {\it Ignore}: a transient where no follow-up observations have been obtained and none are expected to be obtained.
\end{itemize}

\noindent These statuses are broadly descriptive enough to organize, track, and act on most transients as they flow through the follow-up process. However, where these metadata states are insufficient, users can create or assign custom metadata tags (see Section~\ref{sec:queries} and Figure~\ref{fig:tags}) to create new groupings of transients.      

\subsection{Custom Queries and Personal Dashboards}

In addition to screening new transients as they are ingested, users can write and execute custom queries as described in Section~\ref{sec:queries}, and attach them to their personal dashboards as described at the end of Section~\ref{sec:frontend}. These queries can be written to select transients based on interesting properties or trends in the data (e.g., rising light curves, close distances, etc.), and offer ways to define scientific samples for further study.

\subsection{Requesting and Planning Follow-up Observations}
\label{sec:followup}

Each transient within \ysepz\ has an automatically generated transient detail page (see Figure~\ref{fig:detail_page}). Within the detail page, users can navigate to the Follow-up tab which has a series of simple forms to record follow-up requests and to add new observational resources to the system. If a user adds a new resource to the system, it is displayed in a calendar format on the Observing Calendar (see Figure~\ref{fig:obs_calendar}).

To manually request follow-up observations a user can link one of these resources to the transient. Finder charts and airmass plots, which can be used to vet the feasiblity of a follow-up request, are available on the transient's detail page under the Resources tab. Once a request is made, a history of all requests, and their statuses (e.g., {\it Requested}, {\it Successful}, {\it Failed}, etc.), are displayed. All requested transients for an observing night are accessible through the observing calendar, and are displayed with interactive dashboards (see Figure~\ref{fig:obs_night}. Each row of this dashboard contains comments left for the observer and a link back to the transient detail page.

In addition to manual follow-up it is also possible to automatically trigger follow-up observations with \ysepz. As mentioned in Section~\ref{crondataingestion}, API-only CRON jobs can be written to select transients that have been tagged using \ysepz's tagging system (see Section~\ref{sec:queries} and Figure~\ref{fig:tags}), and then inject those transients into third party scripts to trigger queue-based networks like Las Cumbres Observatory's API \citep{Brown2013,Nation2022} or NOIRLab's AEON network \citep{Briceno20}.

\subsection{Uploading Follow-up Data}

Finally, after transient follow-up observations have been successfully obtained a user can upload the data to \ysepz\ via the API. These uploads can be manual, or can be automated by adding an upload stage to an independent data reduction CRON jobs. Once uploaded, the transient detail page can be refreshed to display this new data, and a a user closes the workflow loop shown in Figure~\ref{fig:workflow}.

\section{Case Studies} 
\label{sec:casestudies}

The transient life cycle described above is only one of many possible use cases for \ysepz. Below, we outline several additional collaboration-level workflows that {\ysepz}'s flexible platform enables.

\subsection{Photometric and Spectroscopic Monitoring of Interesting Transients by the UC Santa Cruz Transient Team}

Although transient sample sizes are growing exponentially, some of the most interesting questions related to transients are best answered by observing individual, extraordinary events. The University of California, Santa Cruz Transient Team spectroscopically and photometrically monitors dozens of interesting transients simultaneously. We use \ysepz\ to triage all discovered transients, track interesting objects, and schedule follow-up observations. Here, we describe in detail how The University of California Santa Cruz Transient Team uses the triage tags outlined in Section \ref{sec:triage}.

After a transient discovery is announced on TNS, \ysepz\ will ingest the basic information for this transient. Initially, \ysepz\ labels its transient status as {\it New}, and it will appear on the dashboard under this category.  Some transients are immediately deemed uninteresting (e.g., old discoveries, high Galactic reddening, coincident with a known quasar or Galactic star) and their status is changed to {\it Ignore}.  A particularly interesting transient will either immediately have follow-up observations requested, at which point its status is automatically updated to {\it Followup Requested}, or its status is changed to {\it Interesting}.  The remaining transients have their statuses changed to {\it Watch}.

For the interesting transients, we will usually manually request forced photometry to obtain those data on a timescale faster than the automatic CRON job would provide.  Additional examination of data on the detail page will often result in applying user-defined tags (e.g., ``Young'').

Either through the main dashboard, where one can sort transients based on the time the last data was obtained, or through the personal dashboard, users can easily monitor these interesting events separately from the bulk of active transients. During normal monitoring or in preparation for a classical observing night, we will use ad hoc queries displayed on the personal dashboard and queries from user-defined tags to select targets for follow-up observations. Some ad hoc queries include targets rising quickly and those with new data within a day of discovery.  Useful user-defined tags include ``personal'' tags for individual users (usually marked by their name) and qualitative information such as those objects for which a paper will be written.  We also examine observation requests from previous observing nights to request observations of additional objects.

Once an object has targeted follow-up observations, its status is changed to {\it Following}.  When we no longer expect to obtain more data for an object, we change the status to either {\it Followup Finished} or {\it Needs Template} depending on if a template observation is required.

If no targeted follow-up data is obtained and there is no expectation to obtain data, a transient's status is changed to {\it Ignore}.  Changing the status not only reduces the number of objects that we are actively monitoring, but also stops now unnecessary CRON jobs executing for that transient (see Section \ref{sec:crons}).

The bulk of transients have a life cycle where their status changes from {\it New} to {\it Watch} to {\it Ignore}; this occurs for thousands of transients each year and requires tools to reduce the workload.  Using a query (Section \ref{sec:queries}), we select all transients with a {\it Watch} status and where there has been no new data in three weeks.  These objects are good candidates for changing their status to {\it Ignore} without examining each object in detail.  We have a separate query to look at the subset of those objects that might have interesting aspects such as being particularly bright at the time of the last observation.  After manually examining the potentially interesting objects (and possibly updating their status to Interesting), we can change all remaining object from the initial query to {\it Ignore} in bulk.

\subsection{Survey Planning and Operations with YSE}
\label{sec:YSE_casestudy}

As introduced Section~\ref{crondataingestion}, YSE is a time-domain survey that combines proprietary Pan-STARRS and other public imaging data, along with significant follow-up resources, to survey a large fraction of the high Galactic latitude Northern sky. For an overview of the survey's goals and operations, see \citet{Jones21:yse} and \citet{Aleo2022}.

 \ysepz\ is an integral part of YSE operations, supporting the survey's transient discovery process and organizing its transient follow-up observations.  YSE relies on complex interfaces between reduced data products \citep[via the Image Processing Pipeline (IPP);][]{Magnier20}, initial transient vetting \citep[via the Transient Science Server;][]{Smith2020}, and data synthesis via \ysepz. Critically, \ysepz\ enables a ``secondary'' vetting of potential YSE SN discoveries. The YSE survey cannot easily rule out whether a possible transient is a moving object, such as an asteroid, because YSE has just five minutes between its observations in two different filters. To mitigate this, YSE uploads unverified transients into \ysepz\ to cross-match against other public surveys. 

Vetting these candidates is made easy through \ysepz\ because observers can search both publicly reported transients at the possible SN location, as well as trigger jobs to download ZTF or ATLAS forced photometry. Furthermore, because \ysepz\ collates a variety of archival data, such as host galaxy redshifts, team members can use queries to identify an approximate luminosity for a given transient and help the team decide if the transient is likely real and/or young. \ysepz\ also uses its record of YSE fields and YSE observations to see if publicly reported transients have YSE imaging at their location and then queries forced photometry at that location via the Pan-STARRS IPP.  IPP forced photometry is then added to the YSE database, and can result in recent non-detections that alert the team that a transient is young. By combining these public, private, and archival data streams, \ysepz\ is perfect for this secondary transient vetting, confirming new transient discoveries, and optimizing follow-up decisions. 

\ysepz\ is also used to schedule and monitor the status of every Pan-STARRS observation for YSE.  \ysepz\ stores each Pan-STARRS observation in a {\tt Survey~Observation} model, which includes the status of each observation (i.e., if it was successfully observed) and important characteristics of each observation (e.g., seeing, airmass, etc.). Team members access these metadata from the API to ensure that future observations meet the desired cadence, filter choice, and moon avoidance criteria, while allowing individual survey images to be inspected for photometric quality. The \ysepz\ data model also contains survey-specific objects to model YSE's Survey Minimum Schedulable Blocks (SMSBs), which are logical groupings of survey field centers. These MSBs enable YSE observers to dynamically manipulate and schedule groups of survey fields, e.g., to image nearby SNe or galaxies of interest, through an interactive tool within \ysepz.

\subsection{Query Driven Follow-up Observations with the Keck Infrared Transient Survey}

The Keck Infrared Transient Survey (KITS) is a NASA Key Strategic Mission Support program (Programs 2022A\_N125, 2022B\_N169, 2023A\_N040; PI Foley) to usethe Near-InfraRed Echellette Spectrometer \citep[NIRES;][]{wilson2004} on the 10-m Keck 2 telescope to obtain, and subsequently make publicly accessible, near-infrared (NIR) spectra of all types of astrophysical transients (Tinyanont et al.\ in prep.). Specifically, KITS emphasizes observations of rare transients, as well as obtaining spectra of more common transients at epochs with poor existing NIR spectroscopic data, which will be crucial to interpreting \textit{James Webb Space Telescope} \citep[{\it JWST};][]{Gardner2006} data of transients at higher redshift. These data will also play a key role in planning future time-domain surveys with the upcoming \textit{Nancy Grace Roman Space Telescope} \citep{Spergel2015}.

To accomplish these goals, KITS selects targets with three different criteria powered by queries written with \ysepz's Query Explorer: (1) a survey of SNe Ia at phases with poor NIR spectroscopic observations and/or with large host extinction, (2) a magnitude-limited survey of all transients with $r < 17$~mag, and (3) a volume-limited survey of transients with $z < 0.01$. For examples of these queries, see Appendix~\ref{sec:example_queries}.

In addition to defining samples of transients to observe, the KITS team tags every KITS object with a custom tag. This tag makes querying \ysepz\ for all KITS objects simple, and facilitates tracking the progress of the overall survey, generating survey statistics, producing analysis plots, and planning which are observations are needed to maximize the science goals of the survey.

\subsection{Making Decisions with Archival, Meta, and Astronomical Data Sources with the Swope Supernova Survey}\label{sec:swopesnsurvey}

The Swope Supernova Survey is a time-domain survey that uses the 1-m Swope telescope at Las Campanas Observatory to obtain \emph{uBVgri} follow-up imaging of SNe and other astrophysical transients \citep[][]{Kilpatrick2018}. A chief aim of the SSS is to produce a sample of low-redshift, cosmologically useful Type Ia SNe akin to the Foundation Supernova Survey \citep{Foley2018}. \ysepz\ is used to check if the host of the transient is nearby, whether that host has coincident {\it Chandra} and {\it HST} imaging, and ZTF and ATLAS forced photometry can constrain whether the transient is young. For transients that are assumed to be SN Ias, the {\tt SALT3} function on the transient detail page (see Section~\ref{sec:valued-added services}) can assess important SN properties like phase and luminosity.

Since 2017, SSS has also searched for the electromagnetic (EM) counterparts to gravitational wave (GW) sources, as a part of the One-Meter Two-Hemispheres \citep[][]{Coulter2017, Kilpatrick17} and the Gravity Collective \citep[][]{Kilpatrick21} collaborations. During a search, astronomers do not know when and where to search for EM counterparts beforehand; this information is published by the LIGO-Virgo-KAGRA Collaboration in the form of localization maps. These maps can range in size from tens to thousands of square degrees on the sky \citep{Abbott20}, and depending on the area, subsequent searches can identify hundreds to thousands of transient candidates a night.

It is crucial to locate EM counterparts to GW sources as quickly as possible to understand their physics because they change on fast timescales \citep[][]{Arcavi2018}. This urgency means that most candidates can not be vetted before they are reported, so they are reported through messaging networks like the Gamma-Ray Coordinates Network \citep{GCN} and the Astronomer's Telegram \citep{AstronomersTelegram} instead of the TNS. \ysepz's archival information is key in committing follow-up resources to these candidates. \ysepz\ can be queried to find spatially coincident preexisting transients, candidates can be matched to a host galaxies to estimate their luminosity distance, and forced photometry can reveal pre-GW event variability. If no viable counterparts are discovered, spatial and temporal queries can retrieve all reported candidates (and their data) potentially associated with GW event to rule them out, allowing for search data to be interpreted as upper limits on counterpart models.

\section{\ysepz\ and NASA's Open-Source Science Initiative}
\label{sec:discussion}

Astronomy faces challenges beyond those solely posed by exponentially increasing data volumes. Large transient surveys and their associated software infrastructure (e.g., Alert brokers, TOMs, value-added services) are often sprawling, complex, and difficult to install and run. To ensure that large collaborations run efficiently, that effort is not duplicated, and to broaden collaboration within the scientific community, the {\it way} that software is built is increasingly as important as the software itself. In the face of these challenges, NASA has made a long-term commitment to the Open-Source Science Initiative (OSSI)\footnote{\url{https://science.nasa.gov/open-science-overview}.} with the aim of implementing NASA’s Strategy for Data Management and Computing for Groundbreaking Science\footnote{\url{https://science.nasa.gov/science-red/s3fs-public/atoms/files/SDMWG\%20Strategy_Final.pdf}}. At the heart of the OSSI are four core principles: transparency, accessibility, inclusivity, and reproducibility. 

While {\ysepz}\ was conceived and the initial framework was built before NASA had adopted the OSSI, our initial goals, structures, and development approaches were aligned with its principles, and we now use the OSSI as a guide for our work moving forward. For instance, {\ysepz}\ is transparent; its code base is open source and developed on Github\footnote{\url{https://github.com/davecoulter}.} under a GNU General Public License v3.0. {\ysepz}\ is inclusive; extensive documentation describing how to use the web interface and API, and how to install and develop against the application is hosted by {\tt readthedocs}\footnote{\url{https://yse-pz.readthedocs.io/en/latest/}.}. {\ysepz}\ is accessible and reproducible; its environment has been virtualized and the orchestration between its required services is scripted using {\tt Docker} \citep{Merkel2014}. Enabled by {\tt Docker}, all of {\ysepz}'s required components, i.e., its database, operating systems, standard and customized packages, etc., are packaged together in a way to make the application installable with only a handful of commands\footnote{See \url{https://yse-pz.readthedocs.io/en/latest/install.html} for instructions on how to run \tt{YSE-PZ} using Docker.}.

Beyond the portability that virtualization affords, having {\ysepz}\ constructed this way removes the installation and deployment barrier that can stifle new developers from contributing to the project. To increase the transparency and reproducibility of all contributions to \ysepz, code is merged through pull requests that include both peer reviews and several automatic and manual tests. Promoting code this way builds confidence that code is of high quality, does what it is intended to do, protects our production environment from preventable bugs, and enhances the application's stability. 

\section{Conclusions}
\label{sec:conclusions}

In this work we describe the features and implementation of \ysepz, a transient survey management platform. We highlight our model of the machine-human workflow used to accumulate data on scientifically interesting transients, and discuss how the design and development operations used to develop \ysepz\ are aligned with the OSSI.

There are currently more than a dozen public, active data sources for transient data.  \ysepz\ leverages this public transient survey data by combining sources such as ZTF, ATLAS, ASAS-SN, and {\it Gaia} that make the entire photometric transient light curve public, with other surveys that publish the first photometric epochs or provide public transient spectra.  \ysepz\ then combines these data with private transient data obtained by individual transient science teams.

\ysepz\ has a flexible data model for ingesting these data from multiple data streams.  Data can be easily displayed, queried or downloaded, and the database itself can be queried to search for transients that display certain photometric or spectroscopic behavior.  Archival data and contextual metadata is combined as part of the data model to allow a holistic picture of each transient. 

\ysepz\ also addresses the challenge of optimizing follow-up resources for a given transient science case.  Transient science goals require incorporating considerations such as the brightness of the transient versus the size of the follow-up telescope, the latency of follow-up observations compared to how fast the transient is evolving, the wavelength/frequency range at which the transient must be observed, and the priority of the observations versus the time remaining in a given program.

\ysepz\ addresses these issues by storing each available observing resource and observing night in the database, and listing all facilities in a calendar page.  Users request transients for follow-up observations, with comments that motivate the follow-up request, and the list of transients requested for each resource and observing night are listed on an observing night's detail page.

Although \ysepz's design is effective for the data volume from current ongoing surveys, the next generation of surveys will present new challenges for our transient management workflows. Current data streams are small enough to allow the human-machine workflow shown in Figure~\ref{fig:workflow} to select individual scientifically interesting transients to be followed (see Appendix~\ref{s:pubs} for a list of publications enabled by \ysepz) and to build scientifically useful samples of transients \citep[e.g.,][]{Aleo2022}. However, for next-generation, high-volume transient surveys such as LSST which will find $\sim$10$^{5}$ bone fide transients a year --- at least a 6-fold increase in our current data ingestion rate --- further workflow automation will become increasingly necessary. Despite this, humans will always be more flexible than static code routines and may always be required to manually request follow-up observations on high-value resources; therefore, this next generation of infrastructure needs to do more than simply remove ``human bottlenecks'' and instead empower humans to make effective decisions in real-time. 

Realizing the scientific potential of these data will require a corresponding investment in infrastructure tools. \ysepz\ is poised to meet these challenges by incorporating new technologies that will reconcile automation with human-centered decision-making processes. New messaging protocols like Apache Kafka\footnote{\url{https://kafka.apache.org/}} are moving the astronomical community toward a ``publication-subscription'' model of information ingestion; \ysepz\ will adapt and enable users to subscribe to alert streams from any astronomical data broker with customizable filters. \ysepz\ will also continue to take advantage of machine-learning advances in automatic transient classification \citep[e.g.,][]{Boone2021, Burhanudin2022}, and could be combined with citizen science platforms \citep[e.g.,][]{Christy2022,Zevin2017} to further empower, or even {\it optimize} \citep[e.g.,][]{Walmsley2020,Wright2017}, robust human-in-the-loop decision making at scale. 

\begin{acknowledgments}
The UCSC transients team is supported in part by NASA grants NNG17PX03C, 80NSSC18K0303, 80NSSC19K0113, 80NSSC19K1386, 80NSSC20K0953, 80NSSC21K2076, 80NSSC22K1513, 80NSSC22K1518, and 80NSSC23K0301; NSF grants AST--1720756, AST--1815935, and AST--1911206; grants associated with {\it Hubble Space Telescope} programs DD--14925, DD--15600, GO--15876, GO--16238, SNAP--16239, GO--16690, SNAP--16691, and GO--17128; the Gordon \& Betty Moore Foundation; the Heising-Simons Foundation, and by fellowships from the Alfred P.\ Sloan Foundation and the David and Lucile Packard Foundation to R.J.F.
D.A.C.\ acknowledges support from the National Science Foundation Graduate Research Fellowship under Grant DGE–1339067. D.O.J.\ and P.M.\ acknowledge support from NASA grant 80NSSC21K0834.   A.G.\ is supported by the National Science Foundation Graduate Research Fellowship Program under Grant No. DGE–1746047.  C.D.K.\ is partially supported by a CIERA postdoctoral fellowship.  M.R.S.\ is supported by the STScI Postdoctoral Fellowship. This research has made use of the NASA/IPAC Extragalactic Database (NED),
which is operated by the Jet Propulsion Laboratory, California Institute of Technology,
under contract with the National Aeronautics and Space Administration. We also acknowledge the use of public data from the {\it Swift} data archive.
\end{acknowledgments}

\vspace{5mm}

\software{ Docker \citep{Merkel2014}, {\tt NumPy} \citep{Harris20:numpy}, {\tt Bokeh} \citep{Bokeh18:bokeh}, {\tt Astropy} \citep{Astropy13:paperI, Astropy18:paperII}, 
{\tt Django} (\url{https://www.djangoproject.com}),
{\tt sncosmo} \citep{barbary21}, {\tt MySQL} (\url{https://www.mysql.com/}, \ysepz\ \citep{Coulter2022}, {\tt GHOST} \citep{Gagliano21}, {\tt healpy} \citep{Zonca2019,Gorski2005}, {\tt abseil-py} (\url{https://github.com/abseil/abseil-py}), {\tt ALeRCE} \citep{Forster2021}, {\tt antares-client} (\url{https://gitlab.com/nsf-noirlab/csdc/antares/client}), {\tt asn1crypto}     (\url{https://github.com/wbond/asn1crypto}), {\tt astroplan} \citep{astroplan2018}, {\tt Astroquery} \citep{Ginsburg2019}, 
{\tt  astunparse} (\url{https://github.com/simonpercivall/astunparse}), {\tt attrs} (\url{https://github.com/python-attrs/attrs}), {\tt backcall} (\url{https://github.com/takluyver/backcall}), {\tt beautifulsoup} (\url{https://www.crummy.com/software/BeautifulSoup/}), {\tt bson} (\url{https://github.com/py-bson/bson}), {\tt cachetools}(\url{https://github.com/tkem/cachetools/}), {\tt casjobs} (\url{https://github.com/dfm/casjobs}), {\tt certifi}(\url{https://github.com/certifi/python-certifi}),
{\tt cffi} (\url{https://cffi.readthedocs.io/en/latest/}), {\tt chardet} (\url{https://github.com/chardet/chardet}), {\tt click} (\url{https://github.com/pallets/click/}), {\tt confluent-kafka} (\url{https://github.com/confluentinc/confluent-kafka-python}), {\tt coreapi} (\url{https://github.com/core-api/python-client/}), {\tt cryptography} (\url{https://github.com/pyca/cryptography}), {\tt matplotlib} \citep{Hunter2007}, {\tt cython} (\url{https://github.com/cython/cython}), {\tt dustmaps} \citep{Green2018}, {\tt extinction} (\url{https://github.com/kbarbary/extinction}), {\tt flatbuffers} (\url{https://github.com/google/flatbuffers}), {\tt gast} (\url{https://github.com/serge-sans-paille/gast/}), {\tt google-auth} (\url{https://github.com/googleapis/google-auth-library-python}), {\tt google-pasta} (\url{https://github.com/google/pasta}), {\tt grpc} (\url{https://github.com/grpc/grpc})
{\tt hdf5/h5py} \citep{hdf5}, {\tt html5lib} (\url{https://github.com/html5lib/html5lib-python}), {\tt httpie} (\url{https://github.com/httpie/httpie}), {\tt inda} (\url{https://github.com/kjd/idna}), {\tt imbalanced-learn} (\url{https://github.com/scikit-learn-contrib/imbalanced-learn}), {\tt iminuit} \citep{iminuit}, {\tt importlib\_metadata} (\url{https://github.com/python/importlib_metadata}) {\tt iniconfig} (\url{https://github.com/pytest-dev/iniconfig}), {\tt ipython} (\url{https://github.com/ipython/ipython}), {\tt itypes} (\url{https://github.com/PavanTatikonda/itypes}), {\tt jedi} (\url{https://github.com/davidhalter/jedi}), {\tt jeepney} (\url{https://gitlab.com/takluyver/jeepney}), {\tt jinja} (\url{https://github.com/pallets/jinja/}), {\tt joblib} ({\url{https://github.com/joblib/joblib}}, {\tt joypy} ({\url{https://github.com/leotac/joypy}}), {\tt keras} \citep{Chollet2018}, {\tt keyring} (\url{https://github.com/jaraco/keyring}), {\tt kiwisolver} (\url{https://github.com/nucleic/kiwi}), {\tt lxml} (\url{https://lxml.de/}), {\tt markdown} (\url{https://github.com/Python-Markdown/markdown}), {\tt markupsafe} (\url{https://github.com/pallets/markupsafe/}), {\tt marshmallow} (\url{https://github.com/marshmallow-code/marshmallow}), {\tt marshmallow-jsonapi}, (\url{https://github.com/marshmallow-code/marshmallow-jsonapi}), {\tt mimeparse} (\url{https://code.google.com/archive/p/mimeparse/}), {\tt mpld3} (\url{https://github.com/mpld3/mpld3}), {\tt mysqlclient} (\url{https://github.com/PyMySQL/mysqlclient}), {\tt oauthlib} (\url{https://github.com/oauthlib/oauthlib}), {\tt opt\_einsum} \citep{Smith2018}, {\tt optional\_django} (\url{https://github.com/markfinger/optional-django}), {\tt packaging} (\url{https://github.com/pypa/packaging}), {\tt pandas} \citep{reback2020}, {\tt parso} (\url{https://github.com/davidhalter/parso}), {\tt pexpect} (\url{https://github.com/pexpect/pexpect}), {\tt photutils} \citep{bradley2022}, {\tt pickleshare} (\url{https://github.com/pickleshare/pickleshare}), {\tt pillow} \citep{murray2022}, {\tt pluggy} (\url{https://github.com/pytest-dev/pluggy}), {\tt progressbar2} (\url{https://github.com/WoLpH/python-progressbar}), {\tt prompt-toolkit} (\url{https://github.com/prompt-toolkit/python-prompt-toolkit}), {\tt protobuf} (\url{https://github.com/protocolbuffers/protobuf}), {\tt ptyprocess} (\url{https://github.com/pexpect/ptyprocess}), {\tt py} (\url{https://github.com/pytest-dev/py}), {\tt pyans1} (\url{https://github.com/etingof/pyasn1}), {\tt pycparser} (\url{https://github.com/eliben/pycparser}), {\tt pyerfa} \citep{kerkwijk2021}, {\tt pygments} (\url{https://github.com/pygments/pygments}), {\tt pyparsing} (\url{https://github.com/pyparsing/pyparsing/}), {\tt pytest} (\url{https://github.com/pytest-dev/pytest/}), {\tt dateutil} (\url{https://github.com/dateutil/dateutil}), {\tt python-utils} (\url{https://github.com/WoLpH/python-utils}), {\tt pytz} (\url{https://pythonhosted.org/pytz/}), {\tt PyVo} \citep{becker2022}, {\tt pyyaml} (\url{https://github.com/yaml/pyyaml}), {\tt requests} (\url{https://github.com/psf/requests}), {\tt request-mock} (\url{https://github.com/jamielennox/requests-mock}), {\tt requests-oathlib} (\url{https://github.com/requests/requests-oauthlib}), {\tt rfpimp} (\url{https://github.com/parrt/random-forest-importances}), {\tt rsa} (\url{https://github.com/sybrenstuvel/python-rsa}), {\tt sklearn} \citep{scikit-learn}, {\tt SciPy} \citep{2020SciPy-NMeth}, {\tt seaborn} \citep{Waskom2021}, {\tt secretstorage} (\url{https://github.com/mitya57/secretstorage}),{\tt sfdmap} (\url{https://github.com/kbarbary/sfdmap}), {\tt six} (\url{https://github.com/benjaminp/six}), {\tt sqlparse} (\url{https://github.com/andialbrecht/sqlparse}), {\tt tensorflow} \citep{tensorflow2022}, {\tt termcolor} (\url{https://github.com/termcolor/termcolor}), {\tt threadpoolclt} (\url{https://github.com/joblib/threadpoolctl}), {\tt toml} (\url{https://github.com/uiri/toml}), {\tt tornad} (\url{https://github.com/tornadoweb/tornado}), {\tt traitlets} (\url{https://github.com/ipython/traitlets}), {\tt typegaurd} (\url{https://github.com/agronholm/typeguard}), {\tt typing\_extensions} (\url{https://github.com/python/typing_extensions}), {\tt unicodecsv} (\url{https://github.com/jdunck/python-unicodecsv}), {\tt uritemplate} (\url{https://github.com/python-hyper/uritemplate}), {\tt urllib3} (\url{https://github.com/urllib3/urllib3}), {\tt wcwidth} (\url{https://github.com/jquast/wcwidth}), {\tt webencodings} ({\url{https://github.com/gsnedders/python-webencodings}}), {\tt werkzeug} (\url{https://github.com/pallets/werkzeug}), {\tt wget} (\url{https://pypi.org/project/wget/}), {\tt wrapt} (\url{https://github.com/GrahamDumpleton/wrapt}), {\tt zipp} (\url{https://github.com/jaraco/zipp}), {\tt zope.interface} (\url{https://github.com/zopefoundation/zope.interface}), {\tt asgiref} (\url{https://github.com/django/asgiref/}), {\tt zoneinfo} (\url{https://github.com/pganssle/zoneinfo}), {\tt auditlog} (\url{https://github.com/gmware/auditlog}), {\tt gunicorn} (\url{https://github.com/benoitc/gunicorn}), {\tt tendo} (\url{https://github.com/pycontribs/tendo}), {\tt opencv-python} (\url{https://github.com/opencv/opencv-python}), {\tt 
aladin-lite} \citep{Bonnarel2000,Boch2014} {\tt Eazy PhotoZ} \citep{Gagliano21,Aleo2022}}

\appendix

\section{Example database queries}
\label{sec:example_queries}

This section contains example queries which can be run from the {\tt MySQL Explorer} interface by a user and shows how transients of interest can be selected. A more extensive list of query examples can be found in \ysepz's documentation.\footnote{\url{https://yse-pz.readthedocs.io/en/latest/queries.html}}

\subsection{Selecting volume-limited and recently discovered SNe Ia}

The query below selects SNe Ia discovered within the last 356 days with either a transient or host redshift $<0.03$. The names of the transients meeting this criteria are returned in descending order of their discovery date. In the query below {\tt disc\_date} is the transient discovery data, {\tt photo\_z} is the host photometric redshift, and {\tt  TNS\_spec\_class} is the spectroscopic classification from the transient name server.

\begin{verbatim}
SELECT
    DISTINCT transient.name, transient.disc_date
FROM YSE_App_transient transient
LEFT JOIN YSE_App_host host ON host.id = transient.host_id
WHERE
    (transient.redshift OR host.redshift OR host.photo_z) IS NOT NULL
    AND COALESCE(transient.redshift, host.redshift, host.photo_z) <= 0.03
    AND DATEDIFF(curdate(),transient.disc_date) < 365
    AND transient.TNS_spec_class LIKE '%Ia%'
ORDER BY 
    transient.disc_date DESC;
\end{verbatim}

\subsection{Selecting new southern and unclassified transients}

The query below selects the names of new transients which are spectroscopically unclassified ({\tt TNS\_spec\_class is NULL}), in the southern sky ({\tt dec}$<= -30$), have a flux signal-to-noise of 2 ({\tt flux/flux\_err >= 2}), a magnitude error $< 1.0$ ({\tt mag\_err < 1.0}), and the current day to be $>21$ and $>7$ days from the first and latest distinct detections, respectively. The names of the transients are returned in descending order of first detection. 

\begin{verbatim}

WITH TransientStats AS
(
    SELECT 
        t.id,
        t.`name`,
        MIN(pd.obs_date) AS `first_detection`,
        MAX(pd.obs_date) AS `latest_detection`,
        COUNT(pd.obs_date) AS `number_of_detection`
    FROM YSE_App_transient t
    JOIN YSE_App_transientphotometry tp ON tp.transient_id = t.id
    JOIN YSE_App_transientphotdata pd ON pd.photometry_id = tp.id
    WHERE 
        (pd.flux/pd.flux_err) >= 2 AND 
        pd.mag_err < 1.0 AND
        t.TNS_spec_class IS NULL AND
        t.dec <= -30
    GROUP BY
        t.id
)
SELECT 
    `name`
FROM 
    TransientStats ts
WHERE
    TO_DAYS(CURDATE())- TO_DAYS(first_detection) < 21 AND
    TO_DAYS(CURDATE())- TO_DAYS(latest_detection) < 7 AND
    TO_DAYS(latest_detection) - TO_DAYS(first_detection) > 0.01 AND
    number_of_detection > 1
ORDER BY 
    first_detection DESC;
\end{verbatim}

\subsection{Selecting a magnitude-limited sample}

\begin{verbatim}
SELECT  
    DISTINCT t.name,
    t.ra,
    t.`dec`
FROM YSE_App_transient t, YSE_App_transientphotdata pd, YSE_App_transientphotometry p
WHERE  
    pd.photometry_id = p.id AND     
    pd.id = (
        SELECT 
            pd2.id 
        FROM YSE_App_transientphotdata pd2,     YSE_App_transientphotometry p2 
        WHERE 
            pd2.photometry_id = p2.id AND
            p2.transient_id = t.id AND 
            ISNULL(pd2.mag) = False AND 
            pd2.flux/pd2.flux_err > 3
        ORDER BY 
            pd2.mag ASC
         LIMIT 1
     ) AND 
    
  pd.mag < 17 AND 
  t.`dec` > -30 AND 
  (t.name LIKE '202%' OR t.name LIKE '201%') AND 
  DATEDIFF(curdate(),t.disc_date) < 365 AND 
  t.TNS_spec_class is not NULL AND 
  t.TNS_spec_class != 'CV' AND 
  t.TNS_spec_class != 'SN Ia';
\end{verbatim}

\section{Publications enabled by \ysepz}\label{s:pubs}

A non-exhaustive search of the literature reveals that \ysepz\ has enabled the following publications in some capacity: \cite{Fulton2023}, \cite{Angus2022}, \cite{Aleo2022}, \cite{Davis2022}, \cite{Ward2022}, \cite{Kilpatrick2022}, \cite{Pastorello2022}, \cite{Jacobson-Galan2022b}, \cite{Tinyanont2022}, \cite{Dimitriadis2022}, \cite{Gagliano2022}, \cite{Jacobson-Galan2022}, \cite{Dettman2021}, \cite{Kilpatrick21}, \cite{Wang2021}, \cite{Armstrong2021}, \cite{Jencson2021}, \cite{Barna2021}, \cite{Jones21:yse}, \cite{Hinkle2021},  \cite{Holoien2020}, \cite{Hung2020}, \cite{Jacobson-Galan20}, \cite{JacobsonGalan2020b}, \cite{Neustadt2020}, \cite{Dimitriadis19}, \cite{Jones2019}, \cite{Li2019}, \cite{Kilpatrick2018}, \cite{Kilpatrick2018b}, \cite{Tartaglia2018}.
    
\bibliography{bibliography}{}
\bibliographystyle{aasjournal}

\end{document}